\documentclass{aa}  
\usepackage{graphicx}
\usepackage{txfonts}
\usepackage{epsfig}
\usepackage[round]{natbib}

\newcommand{\nc}{\newcommand}
\nc{\Porb}{$P_{\rm orb}$\,}
\nc{\Teff}{$T_{\rm eff}$\,}
\nc{\logg}{log\,$g$\,}
\nc{\kms}{\,${\rm km\,s}^{-1}$\,}
\nc{\Msun}{$M_{\odot}\ $}
\nc{\Mcz}{$M_{CZ}\ $}
\nc{\vsini}{$v \sin i$} 
\nc{\vmic}{$v_{\rm mic}$}
\nc{\vrad}{$v_{\rm rad}$}
\nc{\ALi}{A$_{\rm Li}$\,}
\nc{\Ali}{A$_{\rm Li}$\,}

\begin{document}

\title{A catalog of rotational and radial velocities
for evolved stars. \\IV. Metal--poor stars \thanks{Based on observations 
collected with the Swiss Euler Telescope, the 1.5--m ESO, the 
2.2--m ESO at La Silla Observatory, ESO, Chile (proposals 69.D--0711 
and 072.D--0315(A), and with the Brazilian - ESO time (proposal 2002.E-12)
operated by the Laborat\'orio Nacional de Astrof\'isica/MCT, Brazil.}$^,$\thanks{
Table 2 is also available in electronic form at the CDS via anonymous ftp 
to cdsarc.u-strasbg.fr (130.79.128.5) or via 
http://cdsweb.u-strasbg.fr/cgi-bin/qcat?J/A+A/XXX/YYY}}

\author{ J. R. De Medeiros\inst{1}
   \and J. R. P. Silva\inst{2}
   \and J. D. do Nascimento Jr\inst{1} 
   \and B. L. Canto Martins\inst{1}
   \and \\L. da Silva \inst{1,3}
   \and C. Melo\inst{4,5}
   \and M. Burnet\inst{6}}

\offprints{J. R. De Medeiros}

\institute{Departamento de F\'{\i}sica, Universidade Federal do Rio 
            Grande do Norte, 59072-970 Natal, RN., Brazil\\
	    \email{renan@dfte.ufrn.br}
       \and Universidade Estadual do Rio Grande do Norte,
                 Mossoro, RN, Brasil
       \and Observat\'orio Nacional, RJ., Brazil
       \and European Southern Observatory, La Silla, Chile 
       \and Departamento de Astronom\'{\i}a, Universidad de Chile,
                 Casilla 36-D,  Santiago, Chile
       \and Observatoire de Gen\`eve, Chemin des Maillettes 51,
            CH-1290 Sauverny, Switzerland}

\date{Received date / Accepted date}

\authorrunning{De Medeiros et al.}
\titlerunning{A catalog of rotational velocities}

\begin{abstract} 
{\sf The present paper describes the first results of an
observational program intended to refine and extend the
existing $v\sin i$ measurements of metal--poor stars, with an
emphasis on field evolved stars.The survey was carried out with the 
FEROS and CORALIE spectrometers. For the $v\sin i$ measurements, 
obtained from spectral synthesis, we estimate an uncertainty of 
about 2.0 \kms. Precise rotational velocities \ensuremath{v\sin i} are
presented for a large sample of 100 metal--poor stars, most
of them evolving off the main--sequence. For the large majority of
the stars composing the present sample, rotational
velocities have been measured for the first time.}
\end{abstract}
\maketitle


\begin{figure}[!htp]
\vspace{.2in}
\centerline{\psfig{figure=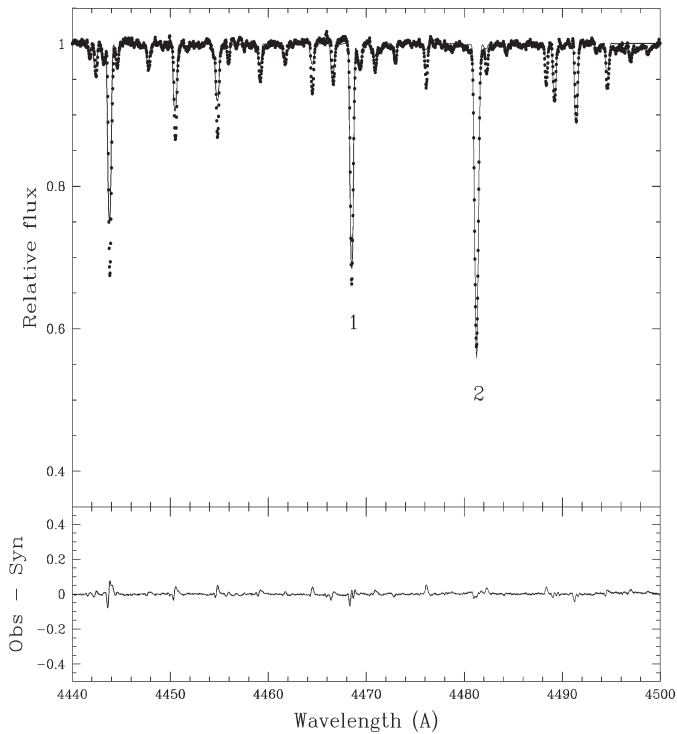,width=3.5truein,height=3.8truein,angle=0}
\hskip 0.1in}
\caption[]{Comparison between the observed and synthetic
spectra for HD 16456 around the 4468\AA\, region.
The solid line and points represent the synthetic and
observed spectra, respectively.
(1) and (2) represent the lines Ti II at 4468.50\AA\,
 and MgII at 4481.20\AA, respectively.}
\label{vsiniF1}
\end{figure}

\section{Introduction}

For the past two decades a large observational effort
has been carried out at the Geneva Observatory,
Switzerland, and at the Department of Physics at Natal,
Brazil, to determine precise rotational velocity for a large
stellar sample at different evolutionary stages and
populations. The major aim behind such an effort is the
understanding of the stellar angular momentum evolution.
For population I evolved stars of luminosity classes IV,
III, II, and Ib, covering the spectral range F to K, a
large set of precise $v\sin i$ measurements was already published 
(De Medeiros and Mayor 1999; De Medeiros et al. 2002, 2004). 
To carry on with these
studies, we turn now to the measurement of $v\sin i$ for
field metal--poor stars.
Since the pioneering works by Peterson (1983, 1985a,b),
there have been an increasing number of observational studies on the
rotational velocity of population II stars ( Peterson et al.
1995; Cohen and McCarthy 1997; Behr et al. 2000a,b; Carney
et al. 2003; Recio--Blanco et al. 2002, 2004), dedicated
mostly to field stars. Nevertheless, in spite of these
solid observational efforts, the data presently available
in the literature are not yet large enough in quantity for a
statistically robust analysis of the rotational behavior
of metal--poor stars in different evolutionary stages.

The present work brings the first results of an observational 
effort dedicated to the study of the rotational velocity of 
metal--poor stars, with an emphasis on field evolved stars. The paper is structured 
as follow: in Sect. 2 the main characteristics of the sample and the
observational procedure are described. The data reduction, spectral
synthesis, and the projected rotational velocity $v\sin i$
measurements are presented in Sect. 3.

\section{The observational program}

For this observing program we have built a preliminary
list of 150 metal--poor stars, listed by Bond (1980) and 
Schuster and Nissen (1988), with Hipparcos parallaxes
and Stromgren photometric indices available
in the literature. Observations were collected with the FEROS
spectrometer mounted on the ESO 1.50--m telescope, and
with the CORALIE spectrometer mounted on the Euler Swiss
1.2--m telescope (Queloz et al. 2000), both at La Silla (Chile). 
Some spectra were obtained in the ESO--LNA Brazil agreement. The FEROS
provides a full wavelength coverage of 3500--9200 Angstroms
over 39 spectral orders at a resolving power R=48,000
(Kaufer and Pasquini 1998), whereas the CORALIE wavelength
coverage is 3875--6820 Angstroms on 68 orders at a
resolving power R=50,000. Both spectrometers are equipped
with a double fiber system, one for the object and the
second for recording the sky or a simultaneous
wavelength calibration.
The exposure times were defined
to produce spectra with a S/N better than 80 at the
spectral region previously chosen to diagnose the
rotation effects on the spectra, namely  lines  at
the range 4440--4500 Angstroms.

\section{The rotational velocity $v\sin i$
computation}

For the determination of the projected rotational velocity 
$v\sin i$, we applied the procedure  by fitting 
the observed spectrum with a synthetic one. 
An essential step in this analysis is the definition of 
suitable spectral lines to measure the $v\sin i$.
The lines that are the best candidates for use in the determination of 
rotation are presented in the Fig. 1. For the present spectral 
analysis we have used the lines listed in Table 1.\\

\begin{table}[!htp]
\begin{center}
\caption{List of the spectral lines used (when possible) 
for the \ensuremath{v\sin i}\ measurement.}
\label{raies_vsini}
\begin{tabular}{ll}
\hline \hline
 Wavelength & Element\cr
  \multicolumn{1}{c}{\qquad(\AA)} & \cr
\hline
4466.55 & \ion{Fe}{i}  \cr
4468.50 & \ion{Ti}{ii} \cr
4481.20 & \ion{Mg}{ii} \cr
4488.33 & \ion{Ti}{ii} \cr
4489.18 & \ion{Fe}{ii} \cr
4491.40 & \ion{Fe}{ii} \cr
\hline \hline
\end{tabular}
\end{center}
\end{table}

\begin{table}[!htp]
\centering
\caption{Rotational velocities for metal--poor stars. The remarks {\it f} and {\it c} 
refer to \vsini\, data from FEROS and CORALIE observations, respectively, {\it V}
to variable stars and {\it SB} for known spectroscopic binary stars.}
\label{results}
\begin{tabular}{lrccc|c}
\hline \hline
Object & $v\sin i$& \Teff & $\log{g}$&  [Fe/H]  & Remarks\\
& (\kms) & (K) &  &   & \\
\hline

        HD        97                &        4.2        &        4953        &        3.1        &$        -1.4$        &        f                \\
        HD        5426                &        5.2        &        5114        &        2.8        &$        -2.1$        &        f                \\
        HD        6268                &        7.5        &        4800        &        0.8        &$        -2.5$        &         f               \\
        HD        13780                &        10.0        &        7930        &        3.1        &        $-1.5$                        &f\\        
        HD        16456                &        15.0        &        7700        &        2.8        &        $-1.5$        &                
	       V,f\\
        HD        19445                &        10.0        &        5911        &        4.3        &$        -1.6$                       
	&V,f\\        
        HD        21581                &        5.0        &        4825        &        2.0        &$        -1.7$        &         f               \\
        HD        22879                &        4.7        &        5808        &        4.2        &$        -0.9$        &         f               \\
        HD        23798                &        6.2        &        4566        &        0.8        &$        -2.1$        &         f               \\
        HD        24289                &        6.3        &        5700        &        3.5        &$        -2.2$        &         f               \\
        HD        25704                &        4.8        &        5830        &        4.1        &$        -1.1$        &         f               \\
        HD        26297                &        5.0        &        4500        &        1.2        &$        -1.7$        &         f               \\
        HD        27928                &        4.0        &        5206        &        2.9        &$        -2.0$        &         f               \\
        HD        29574                &        5.5        &        4310        &        0.6        &$        -1.9$        &                 
	      V,f\\
        HD        31943                &        6.0        &        7690        &        3.2        &        $-1.0$        &                 
	 f     \\
        HD        34328                &        5.5        &        5928        &        4.3        &$        -1.7$        &    f                    \\
        HD        36702                &        5.6        &        4485        &        0.8        &$        -2.0$        &     f                   \\        
        HD        44007                &        5.0        &        4850        &        2.0        &$        -1.7$        &      f                  \\        
        HD        45282                &        3.0        &        5477        &        3.3        &$        -1.4$        &       f                 \\        
        HD        46341                &        4.0        &        5683        &        4.2        &$        -0.8$        &        f                \\        
        HD        51754                &        3.5        &        5830        &        4.3        &$        -0.5$        &    c                    \\        
        HD        51929                &        3.0        &        5886        &        3.5        &$        -0.5$        &    c                    \\        
        HD        56274                &        5.0        &        5700        &        4.3        &$        -0.6$        &  f                      \\        
        HD        63077                &        5.0        &        5715        &        4.1        &$        -1.0$        &  f                      \\        
        HD        63598                &        4.5        &        5852        &        4.1        &$        -0.7$        &  f                      \\        
        HD        74721                &        1.0        &        8900        &        3.3        &$        -1.4$        &  f                      \\        
        HD        76932                &        5.0        &        5880        &        4.0        &$        -1.0$        &  f                      \\        
        HD        78913                &        10.0        &        8515        &        3.2        &        $-1.5$        & f                       \\        
        HD        83212                &        6.0        &        4439        &        1.4        &        $-1.4$&          f      \\                        
        HD        83220                &        8.0        &        6546        &        4.2        &        $-0.7$&        c        \\                        
        HD        84903                &        6.0        &        4700        &        3.3        &        $-1.4$        &                        c        \\
        HD        84937                &        5.2        &        6409        &        3.9        &$        -2.2$        &    f                    \\        
        HD        85773                &        7.0        &        4450        &        1.1        &        $-2.0$        &     f                   \\        
        HD        86986                &        13.0        &        7950        &        3.2        &$        -1.8$        &     f                   \\
        HD        93529                &        8.0        &        4840        &        2.4        &$-1.2$        &          f              \\        
        HD        97916                &        10.2        &        6016        &        4.0        &$        -1.1$        &  f                      \\
        HD        99383                &        4.0        &        6143  &        4.2        &$        -1.5$        &             SB,c           \\
        HD        101063                &        5.0        &        5163        &        3.4        &$        -1.1$        &   f                     \\
        HD        103036                &        8.0        &        4375        &        0.8        &$        -1.7$        &   f                     \\
        HD        103545                &        5.7        &        4725        &        1.7        &$        -2.1$        &   f                     \\
        HD        104893                &        6.0        &        4500        &        1.1        &        $-2.2$        &   f                     \\
        HD        106304                &        5.0        &        9747        &        3.5        &        $-1.5$        &   f                     \\
        HD        110184                &        4.0        &        4366        &        0.5        &$        -2.4$        &   f                     \\
        HD        111721                &        5.0        &        4825        &        2.2        &$        -1.5$        &   f                     \\
        HD        111777                &        5.0        &        5693        &        4.4        &$        -0.7$        &     c                   \\
        HD        111980                &        4.0        &        6032        &        3.7        &$        -0.7$        &       c                 \\
        HD        113083                &        4.5        &        5762        &        4.0        &$        -0.9$        &    f                    \\
        HD        117880                &        16.5        &        7880        &        3.3        &$        -1.6$        &   f                     \\
        HD        118055                &        5.0        &        4088        &        0.8        &        $-1.8$&                        c        \\
        HD        121261                &        5.0        &        4210        &        1.0        &        $-1.5$&      f                  \\        
        HD        122563                &        5.0        &        4697        &        1.3        &$        -2.6$        &               
	V,f\\
        HD        122956                &        8.0        &        4575        &        1.1        &$        -1.8$        &  f                      \\
        HD        126238                &        5.0        &        4979        &        2.5        &        $-1.7$        &  f                      \\
        HD        128279                &        5.0        &        5275        &        2.8        &        $-2.0$        &  f                      \\
        HD        130095                &        7.0        &        9000        &        3.3        &$        -1.8$        &   f                     \\
        HD        132475                &        5.0        &        5920        &        3.6        &$        -1.1$        &           V,c             \\
        HD        134169                &        5.2        &        5861        &        3.9        &$        -0.8$        &   f                     \\
\hline \hline
\end{tabular}
\end{table}

\begin{table}[!htp]
\raggedright
{\bf Table2: } Cont.\\
\centering
\label{results}
\begin{tabular}{lrccc|c}
\\ \hline \hline
Object & $v\sin i$& \Teff & $\log{g}$&  [Fe/H]  & Remarks\\
& (\kms) & (K) &  &   & \\
\hline
        HD        136316                &        4.0        &        4998        &        1.1        &$        -1.4$        &         c               \\
        HD        140283                &        5.0        &        5928        &        3.4        &$        -2.0$        &        V,c               \\
        HD        145417                &        5.0        &        4953        &        4.5        &$        -1.2$        &    f                    \\
        HD        145598                &        5.0        &        5525        &        4.4        &$        -0.6$        &    f                    \\
        HD        148704                &        6.2        &        5096        &        4.0        &$        -0.5$        &               
	        SB,f\\ 
        HD        148816                &        5.7        &        5882        &        4.0        &$        -0.7$        &   f                     \\
        HD        149414                &        5.0        &        5437        &        4.4        &$        -1.0$        &   f                     \\
        HD        149996                &        5.0        &        5700        &        3.9        &$        -0.6$        &   f                     \\ 
        HD        158809                &        5.0        &        5450        &        3.8        &$        -0.5$        &    c                    \\ 
HD        159482                        &        5.0        &        5987        &        4.3        &$        -1.0$        &   f                     \\        
HD        160617                        &        6.2        &        6209        &        3.8        &$        -1.7$        &   f                     \\        
HD        161770                        &        5.8        &        5547        &        3.9        &$        -2.0$        &   f                     \\        
HD        163799                        &        5.2        &        5859        &        3.9        &$        -0.9$        &   f                     \\        
HD        163810                        &        6.0        &        5523        &        4.1        &$        -1.1$        &   f                     \\        
HD        165195                        &        5.0        &        4100        &        0.8        &        $-1.9$        &                        V,c        \\
HD        171496                        &        7.0        &        4700        &        1.6        &$        -0.9$        &   f                     \\        
HD        175179                        &        4.4        &        5830        &        3.9        &$        -0.7$        &   f                     \\        
HD        179626                        &        4.0        &        6106        &        3.7        &$        -0.8$        &     c                   \\        
HD        184266                        &        8.5        &        5500        &        2.5        &$        -1.5$        &    f                    \\        
HD        184711                        &        8.2        &        4500        &        0.6        &$        -2.2$        &                
       V,f\\        
HD        189558                        &        5.2        &        5602        &        3.7        &$        -1.1$        &   f                     \\        
HD        192031                        &        5.0        &        5324        &        4.4        &$        -0.8$        &   f                     \\        
HD        199288                        &        5.0        &        5655        &        4.2        &$        -0.6$        &   f                     \\        
HD        199289                        &        4.0        &        5984        &        4.3        &$        -0.8$        &          c \\        
HD        200654                        &        4.0        &        5477        &        3.6        &$        -2.4$        &          c \\        
HD        200973                        &        4.8        &        6453        &        3.9        &$        -0.5$        &    f                    \\        
HD        201099                        &        5.3        &        5912        &        4.0        &$        -0.5$        &    f                    \\        
HD        204543                        &        4.0        &        5365        &        1.2        &$        -2.0$        &         c\\        
HD        206739                        &        5.3        &        4930        &        1.7        &        $-1.6$        &                        c        \\
HD        212038                        &        4.4        &        5076        &        4.5        &$        -0.5$        &     f                   \\
HD        215257                        &        3.0        &        5978        &        4.4        &$        -0.7$        &  c                      \\
HD        219617                        &        6.2        &        5825        &        4.3        &$        -1.5$        &      f                  \\
HD        222434                        &        5.4        &        4477        &        1.1        &$        -1.7$        &      f                  \\
HD        274939                        &        3.0        &        5282        &        3.0        &$        -1.2$        &  c                      \\
BD        $-1^{\circ}$        1792                &        4.6        &        4850        &        2.7        &$        -1.2$        &  f                      \\
BD        $-1^{\circ}$        2582                &        10.0        &        5130        &        2.4        &        $-2.3$        & f                       \\
BD        $-9^{\circ}$        5831                &        3.0        &        5327        &        1.4        &        $-2.0$        &   c                     \\
BD        $-10^{\circ}$        548                &        4.0        &        5706        &        3.0        &$        -1.5$        &   f                     \\
BD        $+3^{\circ}$        740               &        1.5        &        6075        &        3.8        &$        -2.8$        &    f                    \\
BD        $+6^{\circ}$        648                &        6.0        &        4500        &        1.1        &$        -2.1$        &    f                    \\
BD        $+8^{\circ}$        2856                &        7.0        &        4480        &        1.1        &$        -2.3$        & c                       \\
BD        $+10^{\circ}$        2495                &        3.0        &        5027        &        1.4        &$        -2.0$        & c                       \\
CD        $-24^{\circ}$        1782                &        8.0        &        5300        &        2.8        &$        -2.8$        &  f                      \\\hline \hline
\end{tabular}
\end{table}

For the present analysis we used a version of the 
spectrum synthesis code MOOG (Sneden 1973), using the profiles of a few 
strong lines on the MgII 4481\AA\, region for 
synthesis. For the MOOG analysis we adopted the grid of LTE model 
atmospheres computed by Kurucz \& Bell (1995). Required models were 
interpolated on the grid.
For each star we have selected the synthetic 
spectrum for use as a template that
best matches the derived temperature, gravity, 
metallicity, and broadening mechanisms
(macroturbulence, instrumental, and $v\sin i$). The metalicity was 
determined making use of the calibrations obtained by 
Schuster \& Nissen (1989) and Bond (1980), using the 
Str\"omgren photometry.
The error on the metallicity was estimated taking
into account the uncertainty of about 0.04dex from the
photometry and associated uncertainties, combined with the
uncertainty of 0.16dex from the calibration itself,
leading to a total error of about 0.17dex. In fact this
estimation is true for single stars, since for binary
systems the uncertainty on photometry may increase as a
consequence of light contribution of the components. In
this context, readers should be cautious of the
metallicity listed for binary stars.
For the effective temperature \Teff, 
they were estimated from the
calibrations of Alonso et al. (1996, 1999). The error in this parameter was
derived from the error due to these calibrations and also the error due to
the photometry, resulting in a total error of 111K for
our stars. For the large majority of the stars, the gravity was measured from a relation between \Teff,
visual magnitude, parallax, bolometric correction BC (measured from a 
calibration of Alonso et al. 1995), and mass (obtained from the evolutionary 
tracks of Girardi et al. 1996 and Girardi 2000). The maximum error measured 
for the gravity, 0.15dex, is a sum of the individual errors due to those parameters. 
For a few stars the gravity was taken from the literature, namely HD 136316 and HD 204543 (Fran\c{c}ois 1996);
HD 51754, HD 111980, HD 132475, HD 140283, HD 158809, and HD 179626 (Fulbright 2000); BD $-9^{\circ}$ 5831, 
BD $+8^{\circ}$ 2856, and BD $+10^{\circ}$ 2495 (Carney et al. 2003); HD 99383 and HD 199289 (Nissen et al. 1997); 
HD 51929 and HD 200654 (Axer et al. 1994); HD 215257 (Edvardsson et al. 1993); HD 274939 (Gratton 1994); 
HD 83220 (Nissen and Schuster 1997); and HD 111777 (Jonsell et al. 2005).
The total error on the projected rotational velocity value 
has been estimated 
by computing the quadratic sum of errors induced by errors on individual 
parameters (\Teff, $\log g$, and [Fe/H]). Considering our errors on fundamental 
parameters, $v\sin i$ has an accuracy of about 2.6 \kms.\\

An external comparison with $v\sin i$ measurements of stars
in common with Carney et al. (2003), Behr (2003), and Peterson (1983)
confirms the good accuracy of our $v\sin i$ data.
The rms of rotational velocity differences, from data
listed in Tables 3, 4, and 5 for comparative purposes, is 1.4 \kms, 
indicating that the present $v\sin i$ measurements are as
accurate as the values obtained by these
authors. In fact, this excellent agreement between our $v\sin i$
data and those from the refered authors is more meaningful if we consider
that different approaches have been applied in the $v\sin i$ determinations.
For Carney et al. (2003) and Berh (2003), synthetic spectral analysis 
was applied by adjusting rotational broadening 
of absorption lines to a synthetic template, whereas Peterson (1983) 
made use of spectral synthesis and Fourier Transform techniques.

\begin{table}[!htp]
\begin{center}
\caption{Rotational velocities $v\sin i_{Our}$ estimated in the present work
and by Carney et al. (2003) $v\sin i_{Carney}$.}
\label{raies_vsini}
\begin{tabular}{l|rr}
\hline \hline
Object		&	$v\sin i_{Our}$	&	$v\sin i_{Carney}$	\\
		&	(\kms)		&	(\kms)	\\
\hline
HD 97		&	4.2	&	4.0	\\
HD 5426		&	5.2	&	6.3	\\
HD 21581	&	5.0	&	5.7	\\
HD 23798	&	6.2	&	5.0	\\
HD 27928	&	4.0	&	1.7	\\
HD 36702	&	5.6	&	6.5	\\
HD 45282	&	3.0	&	5.0	\\
HD 83212	&	6.0	&	7.3	\\
HD 93529	&	8.0	&	5.0	\\
HD 103545	&	5.7	&	3.0	\\
HD 111721	&	5.0	&	7.4	\\
HD 118055	&	5.0	&	6.3	\\
HD 121261	&	5.0	&	7.1	\\
HD 122956	&	8.0	&	11.7	\\
HD 171496	&	7.0	&	5.1	\\
HD 184266	&	8.5	&	5.0	\\
HD 206739	&	5.3	&	7.3	\\
HD 222434	&	5.4	&	6.9	\\
BD $-1^{\circ}$ 2582	&	10.0	&	8.0	\\
BD $-9^{\circ}$ 5831	&	3.0	&	5.1	\\
BD $+8^{\circ}$ 2856	&	7.0	&	8.9	\\
BD $+10^{\circ}$ 2495	&	3.0	&	3.8	\\
\hline \hline
\end{tabular}
\end{center}
\end{table}

\begin{table}[!htp]
\begin{center}
\caption{Rotational velocities $v\sin i_{Our}$ estimated in the present work
and by Behr (2003) $v\sin i_{Behr}$.}
\label{raies_vsini}
\begin{tabular}{l|rr}
\hline \hline
Object		&	$v\sin i_{Our}$	&	$v\sin i_{Behr}$	\\
		&	(\kms)		&	(\kms)	\\
\hline
HD 97		&	4.2	&	4.0	\\
HD 74721	&	1.0	&	2.6	\\
HD 86986	&	13.0	&	9.2	\\
HD 117880	&	16.5	&	14.5	\\
HD 161770	&	5.8	&	2.6	\\
HD 184266	&	8.5	&	9.3	\\
BD $+3^{\circ}$ 740 &	1.5	&	6.0	\\
BD $+10^{\circ}$ 2495 &	3.0	&	2.6	\\
\hline \hline
\end{tabular}
\end{center}
\end{table}

\begin{table}[!htp]
\begin{center}
\caption{Rotational velocities $v\sin i_{Our}$ estimated in the present work
and by Peterson (1983) $v\sin i_{Peterson}$.}
\label{raies_vsini}
\begin{tabular}{l|rr}
\hline \hline
Object		&	$v\sin i_{Our}$	&	$v\sin i_{Peterson}$	\\
		&	(\kms)		&	(\kms)	\\
\hline
HD 19445	&	10.0	&	10.0	\\
HD 74721	&	1.0	&	6.0	\\
HD 84937	&	5.2	&	6.0	\\
HD 86986	&	13.0	&	9.0	\\
HD 117880	&	16.5	&	12.0	\\
HD 130095	&	7.0	&	6.0	\\
\hline \hline
\end{tabular}
\end{center}
\end{table}

\section{Contents}

The present study brings projected rotational
velocity $v\sin i$ for 100 metal--poor stars, most of
which were evolving off the main sequence. The main results of 
this part of the program concerning
the rotational velocity for metal--poor stars are listed
in Table 2, where stars appear in order of increasing HD
number, except for those stars presenting only a BD number 
in the literature. The columns mean:\\

1. HD number

2. The measured rotational velocity $v\sin i$

3. Effective temperature

4. Gravity

5. Metallicity

6. Remarks \\

For the stars with a flag {\it f} or {\it c} in the Remarks column, the rotational velocity 
$v\sin i$ was determined
from observations carried out with the FEROS and CORALIE spectrometers, respectively. 
{\it V} in the Remarks column stands for 
variable stars and {\it SB} for 
known spectroscopic binary stars.  \\



  





\acknowledgements{J. R. De Medeiros and his Brazilian
collaborators gratefully acknowledge the Geneva Observatory
for the generous amount of observing time at the Swiss
Euler Telescope provided for the present program. B.L.C.M., J.R.M., 
and J.D.N.,Jr acknowledge the CAPES and CNPq Brazilian agencies. 
This paper took advantage of the useful comments and suggestions 
of the referee, Dr. Bruce W. Carney. This work has been supported 
by continuous grants from the CNPq and FAPERN Brazilian agencies.}

\end{document}